\documentclass{article}
\usepackage[utf8]{inputenc}
\usepackage{csquotes}

\title{A new competition format to avoid collusion in three-team sports tournaments}
\author{José A. Troyano\\School of Computer Engineering\\University of Seville\\{\normalsize troyano@us.es}}
\date{\today}

\begin{document}

\maketitle

\begin{abstract}
The usual FIFA World Cup$\mathrm{^{TM}}$ group stage format is eight groups of four teams for a total of 32 teams.  The top two teams from each group advance to the next round, resulting in a 16-team knockout stage.

The next edition of the World Cup will have 48 teams. Although FIFA initially proposed a configuration of 16 three-team groups, it has ultimately settled on 12 four-team groups. The format chosen, based on four-team groups, has clear organizational disadvantages and even a lack of fairness in the competition. But, in return, there is less risk of collusion in groups of four teams than in groups of three teams.

This study proposes a modification to the three-team group format to reduce the risk of collusion by making the order of the matches not predefined in advance. The results obtained with Monte Carlo simulations show that the new three-team competition format with dynamic match order is comparable, in terms of collusion risk, to the four-team group system commonly used in past World Cups.
\end{abstract}

\section{Motivation}

In 2023, FIFA approved the format of the next 2026 World Cup \cite{fifa2023-eng}. This competition will have 48 teams for the first time, which will be organized into 12 groups of four teams. The top two teams in each group will advance to the next phase, along with the eight best third-placed teams, with the champion being decided by a knockout system among the 32 teams that qualify for the second phase.

FIFA had initially proposed a different competition system, with 16 groups of three teams \cite{fifa2017-eng}. With this distribution in three-team groups, there would be a more natural transition to the second stage, as the top two from each group would complete the 32-team pool for the knockout stage.

There are several advantages of the organization in 16 groups of three teams, especially if the goal is to maintain a tournament of similar dimensions to the one played so far with 32 teams. FIFA itself, in 2017, highlighted these advantages of the 16 three-team group system:  the current duration of the tournament is maintained at 32 days with a moderate increase in total matches (from 64 to 80), the total number of rest days is not reduced, and a maximum of seven matches are guaranteed for the teams that reach the final. 

With the system of 12 groups of four teams, however, the total number of matches increases to 104, which means that the duration of the tournament will be extended to 39 days. This put more physical strain on the players, and clubs will have to release their players to national teams for a longer period of time, making the club competition schedules even tighter.

There is also a competitive asymmetry in the system of 12 groups of four teams. To complete the 32-team pool for the knockout phase, the eight best third-placed teams must be included. This implies that the decision as to which third place teams go through to the next round, and which do not, is made without the possibility of these teams playing against each other. It is also impossible for all potential third-placed teams to play their matches at the same time, so some will play with the advantage of knowing the scores of their competitors to advance to the next round.

Given these drawbacks, and taking into account the initial FIFA proposal of a design of 16 groups of three, why in the end it has opted for a design of 12 groups of four? The answer to this question lies in the probability of collusion, a situation in which there is a result that allows the two teams in a match to qualify for the next stage regardless of what other teams in the group do. 

Specifically, the probability that on the last matchday there will be a result that guarantees that the two teams in a match will advance to the next round is $50.2\%$ for groups of three teams, and $12.2\%$ for groups of four teams. This is because in a three-team group there is always one team that rests on the last matchday, while the other two teams have the advantage of playing their last match knowing the result of all the matches. 

This study relies on a simulation technique, called the Monte Carlo method \cite{metropolis1949monte}, to estimate the probability of collusion with different competition formats, and to show that it is feasible to organize reliable competitions with groups of three teams.

The rest of this study is organized as follows: in section \ref{esquemas} the three different competition formats to be analyzed are presented, in section \ref{experimentacion} the results of the Monte Carlo simulations for the three competition formats are shown, finally in section \ref{conclusiones} the most relevant conclusions of this study are summarized.

\section{Competition formats} \label{esquemas}

The main goal of this study is to evaluate the collusion probabilities of three competition formats. Two of them are conventional systems already used in football competitions in the past, and the last one is a new proposal that considerably reduces the risk of collusion in groups of three teams:

\begin{itemize}
\item Four-team groups with fixed order of matches: this is the most common system in national team competitions. In the all-against-all format six matches are played, and on the last matchday the last two matches are played at the same time so that no team has the advantage of knowing the result of other matches.

\item Three-team groups with fixed order of matches: it is a less common system, but there are precedents in several editions of World Cups. In the all-against-all system, a total of three matches are played. On each matchday, two teams play each other and one team rests, so that on the last matchday there is always one team that depends on the result of the other two. Precisely because of this circumstance the risk of collusion is higher, as there may be results that favour the two teams playing on the last matchday to the detriment of the team that has to rest.

\item Three-team groups with dynamic order of matches: in this new proposal, based on the three-team format, a modification is introduced. Only the first match is known in advance, and the order of the second and third matches is decided on the basis of the result of the first match. The idea is that the team with the worst result in the first match will be the one to play on the final day. This way the bad start is compensated with the advantage of being able to play the last match.
\end{itemize}

Only by applying a dynamic order in the three-team format, the risk of collusion is significantly reduced. In fact, the risk of collusion is almost the same as in the four-team group format.

If there is a tie in the first match a penalty shoot-out would be held, as is done in knockout-based systems, to determine the winner of the first match. In that case the winner of the penalty shoot-out would score two points in the group standings, instead of only one for the draw. 

Thus, for a group of three teams A, B and C, the proposed new format with dynamic match order would look like this, A-B being the first match:

\begin{itemize}
\item If A wins, or there is a tie and A wins the penalty kicks: the order of the remaining matches is A-C, B-C.
\item If B wins, or there is a tie and B wins the penalty kicks: the order of the remaining matches is B-C, A-C.
\end{itemize}

One could even further reduce the risk of collusion by having the two strongest teams play the first match, as proposed in \cite{guyon2020risk}. Since the last match would be played by the {\it a priori} weakest team, and the team that had the worst result in the first match.

\section{Experimentation} \label{experimentacion}
To apply the Monte Carlo method to the collusion risk assessment of the three competition formats, a random match results generator is needed. In this study, the historical results of all matches played in World Cup finals have been used to estimate the probability distribution of the goals scored in a match. And this distribution is the basis for the random results generator.

From the random results generator, as many competition simulations as needed can be produced in order to estimate the risk of collusion for each competition format. Each problem solved with the Monte Carlo method requires a different number of simulations. In the case of the three competition formats analyzed, the results converge after 10,000 iterations, and this is the number of simulations used in the experiments.

The table \ref{dos-equipos} shows, for each competition format, the probabilities that there is a result of the last match that makes the two teams in that match first and second in the group. As can be seen, the three-team format with fixed order has the highest probability. The dynamic schedule variant implies a decrease of 30 percentage points in this probability, demonstrating the effectiveness of this measure with respect to the order of the matches.

\begin{table}[]
\begin{center}
\begin{tabular}{l|c|c|}
 &  probability \\
\hline
 three team-group,fixed order & $50.18\%$   \\
 \hline
 four team-group, fixed order &  $12.23\%$   \\
 \hline
 three team-group, dynamic order &  $21.07\%$   \\
 \hline

\end{tabular}
\caption{\label{dos-equipos}Probability that any result of the last matchday will qualify the two teams of the same game.}
\end{center}
\end{table}

But probabilities of table \ref{dos-equipos} do not really represent the risk of collusion. We have to distinguish between situations in which virtually any match outcome favors both teams, and situations in which very few match outcomes favor both teams. Only in the second case can one speak strictly of collusion, because the two teams have to strive to achieve a result that favors both. To reflect this difference, two situations are defined: high risk of collusion and low risk of collusion. A low risk of collusion is considered when more than $50\%$ of the results of the last match are favorable to both teams. On the other hand, a high risk of collusion is considered when less than $50\%$ of the results of the last match are favorable to both teams. 

\begin{table}[]
\begin{center}
\begin{tabular}{l|c|c|}
 &  low risk & high risk   \\
\hline
 three team-group,fixed order & $12.27\%$ & $18.31\%$  \\
 \hline
 four team-group, fixed order& $4.74\%$ & $5.34\%$ \\
 \hline
 three team-group, dynamic order&  $13.98\%$ & $6.74\%$ \\
 \hline

\end{tabular}
\caption{\label{riesgo}Risk of collusion according to match order.}
\end{center}
\end{table}

The table \ref{riesgo} shows the probabilities of high risk of collusion and low risk of collusion for the three competition systems. As can be seen, in the new proposal of groups of three teams with dynamic order there is little probability of incurring high risk of collusion ($6.74\%$) with a risk very similar to the usually used four-team system ($5.34\%$).

\section{Conclusions} \label{conclusiones}

The results of the Monte Carlo method show that the three-team competition format with dynamic match order is comparable, in terms of risk of collusion, to the four-team group system usually used in past World Cups.

This improvement of the three-team group format, coupled with the choice of the first match with the two strongest teams proposed in 
 \cite{guyon2020risk}, makes it feasible to organize a 48-team World Cup split into 16 three-team groups as FIFA initially intended to do when it first communicated the expansion of the competition from 32 to 48 teams 
 \cite{fifa2017-eng}.

\bibliographystyle{plain}
\bibliography{WorldCup-arXiv}
\end{document}